\documentclass[12pt, onecolumn]{revtex4-2}    

\usepackage[a4paper, left=2.5cm, right=2.5cm,
 top=2.5cm, bottom=2.5cm]{geometry}       

\usepackage[font=small,
labelfont=bf]{caption}                      

\usepackage{graphics,graphicx,epsfig,ulem}	
\usepackage{amsmath} 	
\usepackage{amssymb}
\usepackage{cancel}
\usepackage{subcaption}

\usepackage{etoolbox}                       
\makeatletter
\patchcmd{\frontmatter@RRAP@format}{(}{}{}{}
\patchcmd{\frontmatter@RRAP@format}{)}{}{}{}
\renewcommand\Dated@name{}

\makeatother

\usepackage{fancyhdr}

\def\thesection{\arabic{section}}

\begin{document}

\title{Gauss's Law and a Gravitational Wave}

\author{Olamide Odutola$^1$, Arundhati Dasgupta$^2$}
\affiliation{\normalfont $^1$ Physics and Astronomy, University of Durham, UK.\\ $^2$ Physics and Astronomy, University of Lethbridge, Canada.}

\begin{abstract}              
 We discuss the semi-classical gravitational wave corrections to Gauss's
 law, and obtain an explicit solution for the electromagnetic potential. The Gravitational Wave perturbs the Coulomb potential with a function which propagates to the asymptotics.
\end{abstract}

\maketitle

\section{Introduction}
The discovery of gravitational waves (GWs) has not only opened a window to astrophysical events, but also given us instruments that are sensitive enough to test very weak gravitational phenomena \cite{gw,gw1}. Therefore, new theoretical work acquires meaning and some of the results can be tested, thereby providing evidence for the correctness of the physical theories. In particular, quantum gravity, which has no experimental confirmation as of yet, needs to be tested. Our entire understanding of the visible matter universe is based on the standard model of particle physics, which is quantized. The quantum of the GW - the graviton - is yet to be detected, and theoretical predictions have non-renormalizable quantum interactions. What, therefore, is the story of gravity at tiny length scales? In \cite{coh}, we explored a coherent state for the GW, which would predict semi-classical phenomena at higher length scales than the $10^{-33}$ cm Planck length. Verification of the predictions from the coherent states would provide evidence for an underlying quantum world, which we hope to probe at a later time with more sophisticated instruments and understanding. On this note, we will briefly discuss a modified GW metric that was obtained in \cite{coh} and has a semi-classical correction to it. A similar computation of generalized uncertainty principle correction to a GW detector has appeared in this volume \cite{gup}. We will then solve Gauss's law and find that there are interesting results with the GW metric by itself. What we find can be interpreted as the charge density receiving a correction which is measurable. We will consider a configuration with a point charge at the origin, which places us in the realm of electrostatics. Coulomb's law is valid and gives the electric field, but no magnetic field. We found that if the background of this is not flat space-time but a GW, then there is a non-zero `current' generated. An interesting discussion of a similar phenomenon and its applications can be found in \cite{em3}. Note our work is also different from the example of an oscillatory electron, which is discussed in \cite{elec}. As the change in source is proportional to the GW amplitude, we studied a
‘perturbation’ of Coulomb’s law that is time-dependent and gives rise to a magnetic field. The time-dependent scalar potential does not fall off at infinity but rises with distance. The electric field’s radial component runs to zero at infinity, but the angular components rise as they have the same radial behavior as the potential; this can be measured and we will provide some numerical estimates. We also show that the magnetic potential is generated
in a similar way as the electric potential. A magnetic field will be obtained from this as non-zero, though one that is very weak. In the conclusion, we will discuss the results
in detail.

\section{Gauss's Law and Gravitational Wave}
We solved for Maxwell's equation when investigating the background of a gravitational wave metric, which was corrected using semi-classical coherent states \cite{coh}.
For the Maxwell field, the Lagrangian is:
 \begin{align*}
     \mathcal{L} &= -\frac{\sqrt{-g}}{4}F^{\mu\nu}\ F_{\mu\nu}=-\frac{\sqrt{-g}}{4}F_{\sigma\rho}\ F_{\mu\nu}g^{\sigma\mu}g^{\rho\nu} \\
     &=-\frac{\sqrt{-g}}{4}\left(\partial_\mu A_\nu-\partial_\nu A_\mu\right)\ \left(\partial_\sigma A_\rho-\partial_\rho A_\sigma\right) g^{\sigma \mu} g^{\rho \nu}, 
 \end{align*}
where we assumed a non-trivial metric.

From Euler-Lagrange equations, we obtained the following EoM in presence of a source four-current $j^{\nu}$:
\begin{equation}
    \frac{1}{\sqrt{-g}}\partial_{\mu}(\sqrt{-g}\ F^{\mu\nu})=\frac{1}{\sqrt{-g}}\partial_{\mu}(\sqrt{-g}\ g^{\mu\rho}\ g^{\nu\sigma}\ F_{\rho\sigma})=j^{\nu}.
    \label{eqn:max}
\end{equation}
 In \cite{coh}, which appeared in this volume, we found semi-classical corrections to a GW metric. We used the coherent states in a system of Loop Quantum Gravity (LQG) which is defined on the phase space of LQG canonical variables, i.e., holonomies $h_{e_a}(A)$ and conjugate momenta $ P^I_{e_a} ({\cal E})$.  The holonomy of the gauge connection $A_a$ was obtained from the exponential of a path-ordered integral of the gauge connection over a one-dimensional `edge' $e_a$ which forms links of a graph; meanwhile, the momentum (built from the densitized triads ${\cal E}_a$) was obtained by smearing the triads ${\cal E}_a$ over surfaces $S_{e_a}$ which the edges intersect. In this calculation, we use only the momentum variables, 
 \begin{equation}
     P^I_{e_a}= \int_{S_{e_a}} ^*{\cal E}^I ;\  \  \   P_{e_a}= \sqrt{P^I_{e_a} P^I_{e_a}}. 
 \end{equation}
 and the following relation:
 \begin{equation}
     {\cal E}^a_I {\cal E}^b_I= q q^{ab}.
 \end{equation}
 where ${\cal E}^a_I$ are the density triads and $a,I=1,2,3$ represent the space and internal SU(2) indices respectively; $q_{ab}$ is the three space-metric of the background; and $q$ is its determinant.
 The coherent states are also characterized using a semi-classical parameter $\Tilde{t}\sim l_p^2/\lambda^2$, which is a ratio of the Planck length to the length scale of the system ($\lambda$ being the GW wavelength); it has a range $0<\Tilde{t}<1$. For these purposes, we consider a measurable $\Tilde{t}\sim 10^{-16}$ for a GW with frequency $10^{35} Hz$. This, however, was too high for the observed waves (which had a frequency of ~100 Hz) as their $\Tilde{t}$ was far smaller. For the next generation of detectors which will detect higher frequency waves, see \cite{livrev} for a review.
 
The momenta were generated by smearing the triads over faces of a cube, perpendicular to the edges $e_a$, which are straight lines, along the three axes. This type of discretization is not unique; however, with respect to the continuum limit, it serves the purpose of helping to find a semi-classical correction to the metric, as defined from the operator expectation values of the momentum (a detailed discussion on this topic can be found in \cite{coh}).
The LQG corrected metric of a gravitational wave with polarizations of $h_+=A_+ \cos(\omega (t-z)),\  h_{\times}=A_{\times} \cos(\omega(t-z))$ (as derived in \cite{coh} is as follows:
\begin{equation}
    g_{\mu\nu}=
    \begin{pmatrix}
    -1 & 0 & 0 & 0 \\
    0 & (1+h_{+})(1+2\Tilde{t}f_{x}) & h_{\times}(1+\Tilde{t}f_{x}+\Tilde{t}f_{y}) & 0 \\
    0 &  h_{\times}(1+\Tilde{t}f_{x}+\Tilde{t}f_{y}) & (1-h_{+})(1+2\Tilde{t}f_{y}) & 0 \\
    0 & 0 & 0 & 1+2\Tilde{t}f_{z}
    \end{pmatrix}.
\end{equation}
 
 The determinant of the metric 

simplified to first order in $\Tilde{t}, h_{\times,+}$ yields:

\begin{equation}
    g \approx (1+2\Tilde{t}f_x+2\Tilde{t}f_y+2\Tilde{t}f_z)(h_{\times}^2+h_+^2-1).
\end{equation}

where the semi-classical correction functions in the metric were,
\begin{align*}
    f_i &= f(P_{e_i})\ ,\ f(P) = \frac{1}{P}\left(\frac{1}{P}-\coth(P) \right),\\ \  P_{e_x} &= \frac{\epsilon^2}{\kappa}\left(1+\frac{1}{2}h_+\right)\\ 
    P_{e_y} &= \frac{\epsilon^2}{\kappa}\left(1-\frac{1}{2}h_+\right)\\
    P_{e_z} &= \frac{\epsilon^2}{\kappa}.
\end{align*}
where the $e_i$ refer to straight edges along the x,y,z directions of the three spatial slice of the system; $\epsilon$ represents the graph edge lengths and $\epsilon \rightarrow 0$ gives the continuum geometry; and $\kappa$ is the dimensional gravitational constant, which is expressed in natural units as the Planck length squared.   
We then found the $0^{th}$ component of the Maxwell's equations in a vacuum, i.e., in the presence of no sources. In flat geometry this gives us Gauss's law, but in the background of the new metric, one instead obtains the following:
\begin{eqnarray*}
-\frac{1}{\sqrt{-g}} \partial_i \left(\sqrt{-g} g^{ij} F_{j0}\right)=0\\
\\ \implies g^{xx} \frac{\partial E_x}{\partial x} + g^{yy} \frac{\partial E_y}{\partial y} + g^{zz}\frac{\partial E_z}{\partial z} +g^{xy} \left(\frac{\partial E_y}{\partial x} +\frac{\partial E_x}{\partial y}\right) + g^{zz} E_z \frac{1}{\sqrt{-g}}\frac{\partial \sqrt{-g}}{\partial z}=0.
\end{eqnarray*}
As the metric semi-classical corrections were proportional to the GW, these corrections were found to be functions of $t, z$ (which has been found as such only in \cite{coh}). However, the derivative terms will be proportional to $\tilde t A_+$, which is a product of small quantities; therefore we could neglect them in the first approximation. Thus, we obtained
\begin{equation}
 \vec{\nabla}\cdot\vec{E}=2 \tilde{t} (f_x\frac{\partial E_x}{\partial x} + f_y \frac{\partial E_y}{\partial y} + f_z\frac{\partial E_z}{\partial z}) + h_+\left(\frac{\partial E_x}{\partial x} - \frac{\partial E_y}{\partial y}\right)+ h_{\times}\left(\frac{\partial E_y}{\partial x}+ \frac{\partial E_x}{\partial y}\right). 
 \end{equation}
In the approximation, we wrote the electric field as a zero-eth order field plus a small perturbation, and the RHS of the above equation could be interpreted as a source for the perturbation. The zeroeth order field is a static EM field generated by a point source at the origin. Hence, we obtained
\begin{equation}
    \vec{E} = \frac{1}{4 \pi \epsilon_0} \frac{\hat{r}}{r^2} + \vec{
    {\tilde E}},
    \end{equation}
where we assumed a point source charge at the origin, or at least a charge of 1 Coulomb within a small radius $\epsilon$ (which is where our considerations were outside the radius).
As the source was time dependent, we took the perturbation to be composed of the potentials
\begin{equation}
    \vec{\tilde{E}}= -\vec{\nabla}\Phi + \frac{\partial \vec{A}}{\partial t}.
    \end{equation}
In the Coulomb gauge $\vec{\nabla}\cdot\vec{A}=0$, the following was yielded:
\begin{equation}
    \nabla^2 \Phi(x,y,z,t)= 6 h_{\times} \frac{xy}{r^5} + 3 h_+\frac{(x^2-y^2)}{r^5}.
\end{equation}
which is clearly Poisson's equation with a time dependent source. Seeing as the divergence of the electric field was zero and to first order in the corrections, all of the $f(P_{e_i})$ were found to be equal, we can ignore the semi-classical term (=$2 \tilde{t} f\vec{\nabla}\cdot\vec{E}=0$). A way through which to understand the GW-generated oscillation of the source is to observe that the charge density fluctuates with time, as the volume changes. 

  To simplify the system,  at $\theta=\pi/2$, we solved for the equations. We got as the particular solution, the following:
\begin{equation}
    \Phi(r,t)= \left( -\frac{3 A_+}{4 r}\right) \cos(2\phi) \cos(\omega t).
\end{equation}
Clearly, this potential is different in behaviour to the regular $1/r$ spherical potential of the point charge source at the origin. Here, the $\phi$ dependence makes the potential acquire different signs as it approaches the x and y axes.
    If we write the above equation in spherical coordinates, assuming a form of the potential in spherical harmonics with the same frequency as that of the GW in its time dependence, we obtain
    \begin{equation}
        \Phi(r,\theta,\phi,t)= \sum_{lm} \Phi_{lm}(r,t) Y^m_l(\theta,\phi).
    \end{equation}
which gives, from Gauss's law, the following:
\begin{equation}
    \sum_{l,m} \left[\frac{d}{dr} \left(r^2 \frac{d \Phi_{lm}}{dr}\right)- l(l+1)\Phi_{lm}\right] Y_{lm}(\theta,\phi)= \frac{3 A_+ e^{i \omega(t-z)}}{r}\sin^2\theta\cos(2\phi).
\end{equation}
proaches the x-axis and the y-axis. The solution also propagates in time.

We then assumed that $\Phi_{lm}(r,t)= e^{i\omega t} \Phi_{lm}(r)$. If we keep the plane wave $e^{ikz}$ in the source $(k = \omega)$, then we have to use the spherical wave expansion of the function $e^{i k r ~\cos\theta}$, given by 
\begin{equation}
    e^{i kr ~\cos\theta }= \sum^{\infty}_{l=0} i^l (2l+1) j_l (kr) P_l (\cos\theta).
\end{equation}

Using the partial wave analysis of the above RHS (with the assumption that the EM potential has the same frequency as the GW), a propagating mode was generated as in the case of oscillating sources. We also wrote the equation $\cos(2\phi) = 1/2\left(\exp( 2i \phi) + \exp(- 2i \phi)\right)$. We found that the ODE for $\Phi_{l2}(r)$ was the same as the ODE for $\Phi_{l-2}(r)$; therefore, we dropped the second index and solved for the following equation:
\begin{eqnarray}
 \sum_{l} \left[\frac{d}{dr} \left(r^2 \frac{d \Phi_{l}}{dr}\right)- l(l+1)\Phi_{l}\right] \sqrt{\frac{(2l+1)}{4\pi}\frac{(l-2)!}{(l+2)!}}P^2_{l}(\cos\theta)  &&\nonumber \\ = \frac{3 A_+}{2} \sum_{l'} i^{l'} (2l'+1) \frac{j_{l'}(kr)}{r} P_{l'} (\cos\theta) \sin^2\theta. &&
\end{eqnarray}
The associated Legendre function $P_l^2(\cos(\theta))$ is on the left and the usual Legendre function $P_l(\cos\theta)$ on the right. If we take the orthonormality property of the associated Legendre functions by first multiplying with $P_n^2 (\cos\theta) d(\cos\theta)$ and then integrating both sides of the Equation for $-1\leq \cos\theta\leq 1$, we obtain
\begin{eqnarray}
    \sum_l \left[\frac{d}{dr} \left(r^2 \frac{d \Phi_{l}}{dr}\right)- l(l+1)\Phi_{l}\right] \lambda_{l}\int_{-1}^1 P_l^2(x) P_n^2 (x) dx && \nonumber \\ = \frac{3 A_+}{2} \sum_{l'} i^{l'} (2l'+1) \frac{j_{l'}(kr)}{r} \int_{-1}^1 P_{l'} (x) (1-x^2) P_n^2 (x) dx, &&
\end{eqnarray}
where $\lambda_l$ represents the normalization constant from the $Y_{lm}(\theta,\phi)$. Furthermore, we replaced $\cos\theta$ with $x$ for brevity.
The LHS uses the orthogonality condition; but on the RHS, the integral is difficult to compute. Given the Legendre function recursion equations \cite{grad} and integrals \cite{sama}, we obtained non-zero values for $l=n-2,n, n+2$. Therefore, we found 
\begin{eqnarray}
    \left[\frac{d}{dr} \left(r^2 \frac{d \Phi_{n}(r)}{dr}\right)- n(n+1)\Phi_{n}(r)\right] \frac{2(n+2)!\lambda_n}{(2n+1)(n-2)!}= &&\nonumber\\
    \frac{1}{ r}\left[\Lambda_{n-2} j_{n-2}(kr)  +\Lambda_{n} j_n(kr) +\Lambda_{n+2} j_{n+2}(kr)\right],&&
    \label{eqn:GGW}
\end{eqnarray}
where there were also the following constants: 
\begin{eqnarray}
    \Lambda_{n-2} &= &\frac{3A_+}{2}i^{n-2}\left[\frac{2n(n^2-1)(n+2) }{(2n+1)(2n-1)} \right],\\
    \Lambda_n &=& -\frac{3 A_+}{2} i^n\left[\frac{4n(n+1)(n-1)(n+2)}{(2n-1)(2n+3)} \right], \\
    \Lambda_{n+2} &=& \frac{3 A_+}{2} i^{n+2}\frac{2n(n+1)(n+2)(n-1)}{(2n+1)(2n+3)}.
    \end{eqnarray}

    There were, therefore  three independent $l=n-2,\ n,\ n+2$ partial waves, which give non-zero values for the RHS of the equation and generate the `source' for the EM potential of $n^{th}$ angular mode. We used MAPLE to generate the solution to the above ODE and we found a very elongated formula containing LommelS1 and Hypergeometric functions, which nevertheless gave the RHS particular solution. It must be noted that if we keep the $\Tilde t$ term detailed in the above equation, the particular solution will get corrected with static functions as there are no-time dependent contributions to first order in $\Tilde t$. As mentioned earlier, we ignored the $\Tilde t A_+$ product terms, which are equivalent to second order infinitesimal corrections to Gauss's law.

    The general solution is:
{\small \begin{eqnarray}
\Phi_n(r) &=& A_0 r^n + \frac{B_0}{r^{n+1}} -\frac{k^{3/2}}{\Gamma\left(\frac72+n\right) 2^{n-1/2}}\left(A \frac{(rk)^{n+1}}{32(n+1)(n+1/2)} H([n+1],[2+n,\frac72 +n],-\frac{r^2 k^2}{4}) \right. \nonumber \\&+& B \frac{(kr)^{n-1}(n+\frac52)(n+\frac32)}{8 n(n+\frac12)} H([n],[n+1,n+\frac32], -\frac{r^2 k^2}{4}) \nonumber \\ &+& \left.C\frac{(n-\frac12)(n+\frac32)(n+\frac52)(rk)^{n-3}}{2(n-1)}H([n-1],[n,n-\frac12],-\frac{r^2 k^2}{4})\right) \nonumber \\
&+&\frac{(rk)^n k^{3/2}}{96 n(n + 1)(n + 1/2)(2 + n)}\left[\left(-\frac18(rk) J_{n-1/2}(kr)+ \frac14  (n+\frac12)J_{n+\frac12}(kr)\right)W(n)S_{3/2-n,n+1/2}(kr) \right.\nonumber \\ 
&+&\left.-n(kr) W(n) J_{n+1/2}(kr)S_{1/2-n,3/2+n}(kr) \right]+ \frac{1}{96(n+\frac12)n(n+1)(2+n)}\left[\left(-\frac14 (kr)^{1/2}W(n) \right.\right.\nonumber\\ &-&\left.2An(n+1)(n+\frac32)(n+\frac12)(kr)^{-7/2}+\frac14 (kr)^{-3/2}nW(n)\right)J_{n+1/2}(kr) \nonumber \\ &+& \left.\left(\frac14 (kr)^{-1/2}nW(n)+(kr)^{3/2} \frac18 W(n)+ (kr)^{-5/2}V(n)\right)J_{n-1/2}(kr)\right].
\label{eqn:solution}
 \end{eqnarray}}
 In the above, we have $J_n(x)$ as the Bessel function of the first kind, $S_{n,m}(x)$ as the LommelS1 functions, $H(a,b;c,d,e,x)$, and $H(a;b,c,x)$ as the Hypergeometric functions of the (2,3) and (1,2) type, respectively. In addition, the $\Phi_n$ have the usual partial wave potentials of the form $r^n$ and $r^{-n-1}$, which were solutions of the homogeneous equation. The particular solutions represent the functions generated by the GW-induced oscillations and are propagating EM potentials. There are singularities hidden in the LommelS1 functions for the integer values of $n$ which we regulated. Note we can trust only the solutions for $r\neq 0$; this is justified as we have a semi-classical parameter $\Tilde{t}\approx 0$, and the discretization $\epsilon$ length scale, which provide a minimum length to which geometry can be probed. The constants are
 \begin{eqnarray}
     A &=&\frac{(2n+1)(n-2)!}{2 (n+2)! \lambda_n} \Lambda_{n+2},\\
     B&=&\frac{(2n+1)(n-2)!}{2^ (n+2)! \lambda_n}\Lambda_n,\\
    C&=&  \frac{(2n+1)(n-2)!}{2 (n+2)!\lambda_n}\Lambda_{n-2},\\
    W(A, B, C)&=& \frac43 ( Cn^2) + (-2B + 4C)n + A - 4B + \frac83 ( C),\\
    V(A, B, C)&=& -\frac{2}3 (Cn^2) + (A - C)n + \frac32 A + \frac{2}3 (C).
 \end{eqnarray}
 Note the above results are true only for $n=2$ and higher. 
As the behavior of the functions for general $n$ were difficult to plot, we simply took one representative partial wave and observed the difference from a regular solution. We take $n=3$ and observe the behavior of $\Phi_3(r)$ as $r\rightarrow \infty$. The $\Phi_3(r)$ function has a real component that falls of as $r^{-4}$, obtained from the homogeneous equation solution; and an imaginary component (which was evident from the coefficients on the RHS) which was the particular solution for $n=3$. Additionally, as our ansatz for the potential is of the form $\Phi(r) e^{i\omega t}$, it was not surprising that the solution was complex. We then plotted the function $|\Phi_3(r)|^2$ to examine its asymptotic behavior. We found that, despite putting the particular solution strength to be $10^{-10}$ of the $r^{-4}$ term, the function started increasing after a certain interval. We know $r^{-4}\rightarrow 0$ as $r\rightarrow \infty$, but the presence of a GW reverses the fall off. This behavior persists for higher $n$, thus confirming our claim that the electric potential now extends to the asymptotic region.

\begin{figure}[h]
\centering
\includegraphics[width=0.5\linewidth, height=7cm]{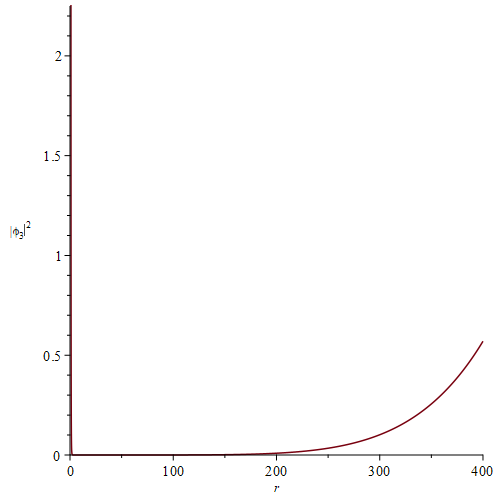}
\caption{The modulus square of the potential for $l=3$}
\label{fig:electric1}
\end{figure}

In general the solutions will be of the form
\begin{equation}
\Phi(r,\theta,\phi,t)= \sum_l \Phi_l(r) \left(P^2_l(\cos\theta) e^{ 2i\phi} + P^{-2}_l(\cos\theta)e^{- 2 i\phi}\right)e^{i \omega t}.
\end{equation}

To obtain the observable function, one must take the real part of the summed solution. 
As shown above in Equation(\ref{eqn:solution}),$\Phi_l(r)$ is composed of solutions to the homogeneous equations of the form $A_l r^l + B_l r^{-(l+1)}$. In addition, for each $l$, there is a particular solution. It is plausible that the sum over $l$ for the particular solution has a finite convergent answer. We tried finding a convergent answer, but the summation was not simple; work is in progress. We instead used a numerical method of summing up the partial waves up to some finite number. We have plotted the particular solution summed up to $l=3...m$, where $m$ is some large number. This evidently represents a truncated GW wave contribution up to the $m+2$ mode in the source, but it is a good-enough approximation to what might be the real system. Therefore, We - in the following -  plotted the plane wave summed up to the $m=50, 100$, as well as showed the corresponding Coulomb potential that was generated by the system. 

We are currently investigating the analytic formula in Equation (\ref{eqn:solution}) and the partial wave summation of the spherical wave solution.  We found that the potential starts growing as observed for the $l=3$ solution of the potential, shown in Figure(\ref{fig:electric1}). We have plotted in 3d the potential plotted for $\phi=0$. This shows that the GW effect on the Coulomb potential was non-trivial and, in principle, was detectable using an electrometer, which is sensitive to the electric potential. This approach will aid in the detection of a GW in a very isolated environment.

\begin{figure}[htbp]
    \begin{subfigure}{0.5\textwidth}
      \includegraphics[width=\textwidth]{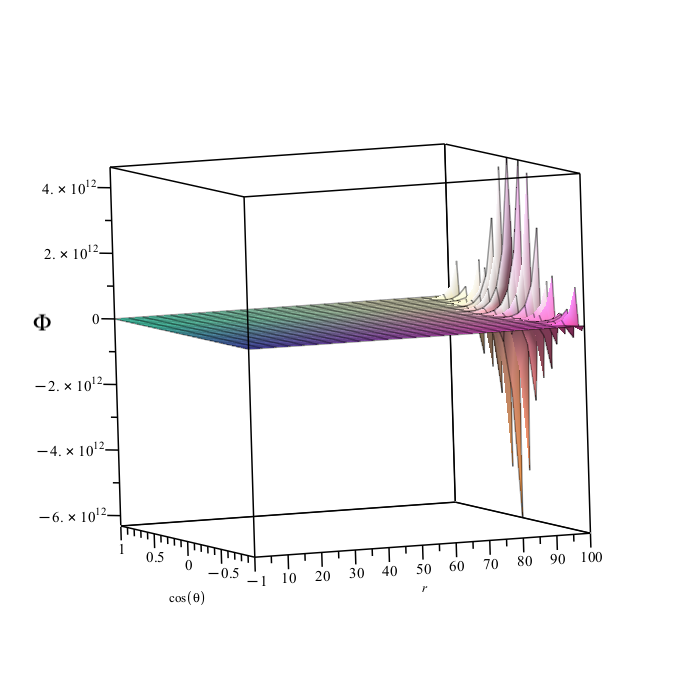}
      \caption{Potential with partial modes summed from $l=3...50$, $\phi=0$.}
      \label{fig:electric2}
    \end{subfigure}
    \hfill
    \begin{subfigure}{0.5\textwidth}
      \includegraphics[width=\textwidth]{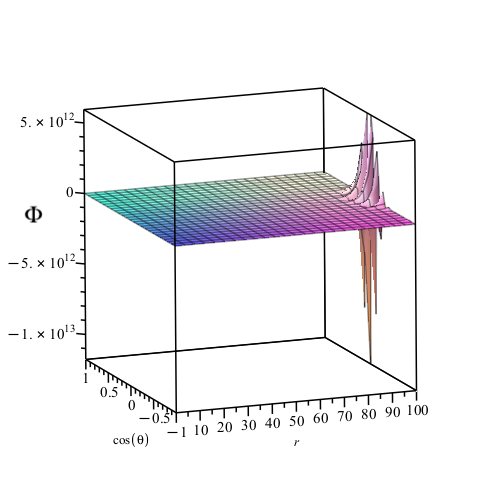}
      \caption{Potential with partial modes summed for $l=3...100$, $\phi=0$.}
      \label{fig:electric3}
    \end{subfigure}
    \hfill
    \caption{Real part of $\Phi(r,\theta,0)$. }
   \end{figure}

\begin{figure}[htbp]
    \begin{subfigure}{0.5\textwidth}
      \includegraphics[width=\textwidth]{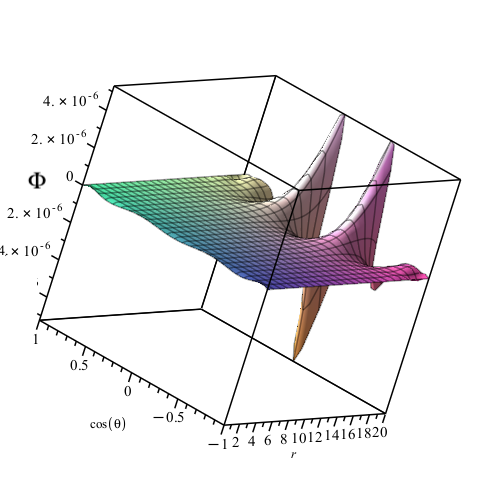}
      \caption{Potential with partial modes summed from $l=3...50$, plotted for $k=1, r=[0,20]$,$\phi=0$.}
      \label{fig:electric123}
    \end{subfigure}
    \hfill
    \begin{subfigure}{0.5\textwidth}
      \includegraphics[width=\textwidth]{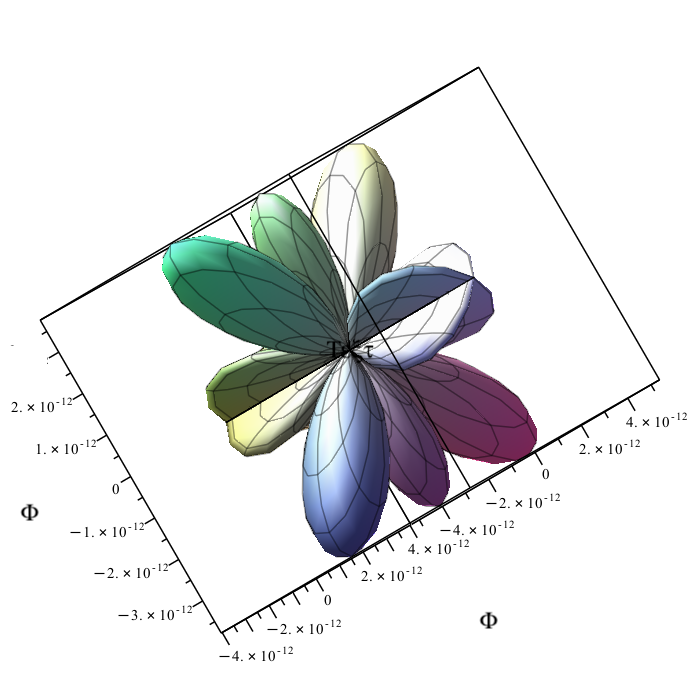}
      \caption{Potential with partial modes summed for $l=3...50$, plotted in $\theta,\phi$, $k=1,r=1$.}
      \label{fig:electric1234}
    \end{subfigure}
    \hfill
    \caption{}
    \label{fig:psi}
\end{figure}

As is evident from the above plots Figures(\ref{fig:electric2},\ref{fig:electric3}) (one for $m=50$, and another for $m=100$), the potential increased as a function of $r$ and the image on the $x=\cos\theta$ axis showed oscillations due to the Legendre function. If one plots the sum over a small interval, then these features are also evident, as shown in Figure(\ref{fig:electric123}). If one plots the potential on the sphere, the oscillations would of course appear as `petals' in a spherical coordinates plot, as shown in Figure(\ref{fig:electric1234}).

The Electric field defined from the above potential was expressed simply as
\begin{equation}
    \vec{E}= -\vec{\nabla}\Phi(r,\theta,\phi)= -\left(\frac{\partial \Phi}{\partial r} \hat{r} + \frac{1}{r}\frac{\partial \Phi}{\partial \theta}\hat{\theta} + \frac{1}{r \sin\theta}\frac{\partial \Phi}{\partial \phi}\hat{\phi}\right).
\end{equation}

The Electric field in the $\hat{r}$ direction has a non-trivial derivative in the radial direction. The derivatives of $\theta$ and $\phi$ acted on the $P_l^2(\cos\theta)$ and the $\cos(2\phi)$. We found $E_r$ from the derivative of the potential function given in Equation(\ref{eqn:solution}); it was also found to be lengthy and involved SturveH functions. Instead of quoting that, we show the graphical representation of the functions in the following (\ref{fig:electric2}, \ref{fig:electric3})for the $l=3$ partial wave only.

\begin{figure}[htbp]
\begin{subfigure}{0.5\textwidth}
	\includegraphics[width=\linewidth, height=5cm]{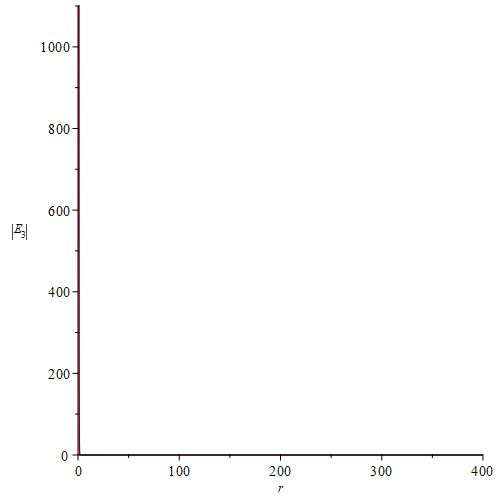}
	\caption{\centering}
\end{subfigure}
\vspace{0.5cm}
\begin{subfigure}{0.5\textwidth}
	\includegraphics[width=\linewidth, height=5cm]{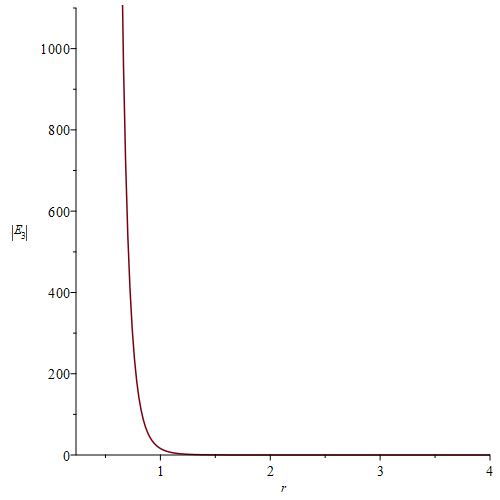}
	\caption{\centering}
\end{subfigure}
\label{fig4}
\caption{Magnitude	of the radial electric field solution for $l=3$. (\textbf{a}) Magnitude of ${E}_r\propto-\partial_r\Phi_3$ for $\phi=0,\ \theta=\pi/4$; $r$ = [0, 400]. (\textbf{b}) The ${E}_r$ field for $\phi=0,\ \theta=\pi/4, r=[0,4]$.}
\end{figure}

\begin{figure}[htbp]
\includegraphics[width=0.5\linewidth, height=6cm]{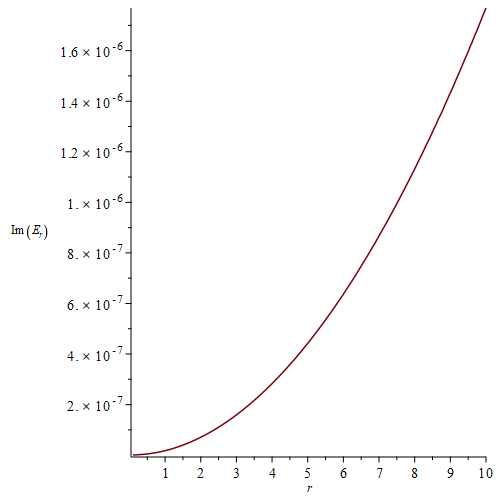}
\caption{The GW-induced electric field radial component for $l=3$, $k$ = 1.}
\label{fig:electric4}
\end{figure}

As evident from the above, the radial component decreased with distance. However, it must be mentioned that the particular part of the solution does show an increase as a function of $r$. As in the potential, we took the ratio of the Coulomb term and GW-induced term as $10^{-10}$. In the event that this ratio was different, the nature of the electric field's radial component will again change.
As shown in Figure (\ref{fig:electric4}), the contribution from the GW-induced electric field increased with $r$. It also remains that there are angular components of the electric field, which are generated due to the GW, and these should be detectable in an electrometer.

\begin{figure}[htbp]
\includegraphics[width=0.6\linewidth, height=8cm]{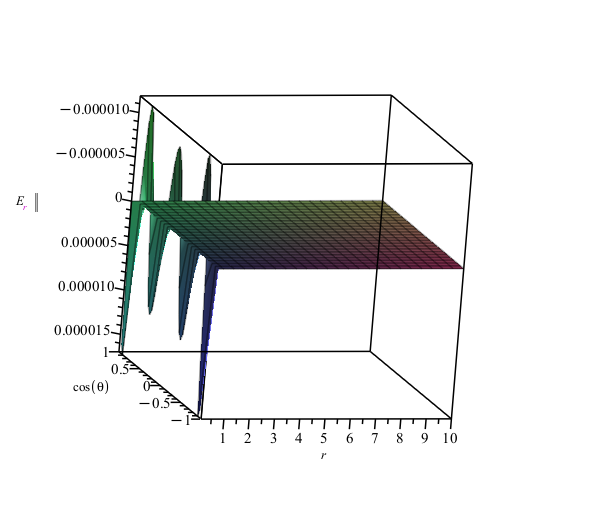}
\caption{The GW-induced electric field radial component for the partial wave summed as $l=3,\ldots,50$, \mbox{$k$ = 1.}}
\label{fig:electric44}
\end{figure}

Next, we also found the electric field's radial component for the summed potential. \[E_r(r,\theta)= -\partial_r\left( \sum_{l=3}^{50} \Phi_l(r) P_l^2(\theta)\right).\] This shows almost similar behaviour as the electric field for $l=3$, the function shows a fall off as a function of $r$. We have plotted  the particular solution or the GW induced electric field, and it is non-trivial, for k=1 as shown in Figure(\ref{fig:electric44}).

Before we end this discussion, the obvious question is whether a calibrated electrometer will detect the above generated fluctuating electric field, and the answer is yes. If we find the potential function at a distance of 10 m from the origin where a $10^{-9}$ Coulomb charge has been placed ($q/4\pi \epsilon_0\sim 1$), and where the GW has a frequency of 10 Hz with an amplitude of $10^{-21}$, then the $E_{\theta}$ component at a fixed angle being proportional to the potential is almost of order 0.1 N/C. Small changes in magnetic fields are detected by SQUIDS\cite{lobo}, we therefore need to discuss the magnetic field generated by the GW.
In the above, we showed how a GW can modify Gauss's law but where our electric field perturbation was time dependent. Therefore, the discussion is incomplete without discussing the magnetic field and studying the vector potential. To obtain the magnetic field, we studied the Maxwell's equations for $\nu=i$, where $i$ is a space component and the current density$j^i=0$, as we are only studying Coulomb's law for a static source in this discussion. We found that the Maxwell's equation is as follows:
\begin{equation}
    -\frac{1}{\sqrt{-g}} \partial_0 \left(\sqrt{-g} g^{ij} F_{0j}\right) + \frac{1}{\sqrt{-g}}\partial_k \left(\sqrt{-g} g^{kl} g^{ij} F_{lj}\right)=0.
\end{equation}
As the magnetic field was initially zero, the contribution to a non-zero magnetic field $\vec{B}$ at first order in the GW amplitude was
\begin{eqnarray}
\left(\partial_z B_y -\partial_y B_z\right) &= & -\partial_0 h_+ E^0_x + \partial_0 {\tilde E}_x ,\\
\left(\partial_x B_z - \partial_z B_x \right) &= &\partial_0 h_+ E^0_y + \partial_0 {\tilde E}_y, \\
(\partial_x B_y - \partial_y B_x) &=& \partial_0 \tilde{E}_z .
\end{eqnarray}
In the above, $E^0_i$ are the components of the Coulomb field and the $\tilde{E}_i$ are the perturbations computed due to the GW. If we use the Lorenz gauge and write the magnetic field in terms of a gauge potential $\vec{B}=\vec{\nabla} \times \vec{A}$, such that $\vec{\nabla} \cdot \vec{A}=0$, one obtains
\begin{eqnarray}
    \nabla^2 A_x & = & -\partial_0 h_+ E^0_x + \partial_0 {\tilde E}_x,\\
    \nabla^2 A_y &=& \partial_0 h_+ E^0_y + \partial_0 {\tilde E}_y, \\
    \nabla^2 A_z &=& \partial_0 \tilde{E}_z .
    \label{eqn:array}
\end{eqnarray}
The above equations can be solved using the same method as the scalar potential solution for Gauss's law.
Thus, apart from modifying Gauss's law, the GW also induces a magnetic field, and this can be calculated. We hope to discuss this in a future work. The fact that a tiny magnetic field was generated is important for detection purposes as small changes in magnetic fields can be found using SQUIDS \cite{lobo}. 

 \section{Conclusion}
 In this short article, we have shown that the GW generates a source for a perturbation of the EM potential, which is time dependent. The solution is complicated in form but was exactly obtained. As GWs were detected, we predicted the corrections to the Coulomb potential of a point source charge, and we hope to find an experimental verification of our results. The semi-classical corrections to the metric described in the paper will also correct Gauss's law in a slightly similar functional form but to the next order in the perturbation. Previously, and in recent years, GW-induced corrections to Maxwell's equations have been studied \cite{em3,emw,lobo,em2,em4,em5}, but our results specifically discussed corrections to a static electric Coulomb potential using partial wave analysis. We also showed how a magnetic field is generated by the GW. We found that, when using numerical values, the GW-induced electric fields propagated and can be almost order 1. The question then is, have we already seen the GW-induced correction to Gauss's law in some detector? To attribute the EM detection to a GW would therefore be the next task.
 \\ \\
 {\bf Acknowledgement:} OO is funded by a MITACS summer fellowship. We are grateful to N. R. Gosala for useful discussions.

\end{document}